\documentclass[aip,graphicx,ams]{revtex4-1}

\usepackage{graphicx}
\usepackage{color}

\draft

\newcommand{\Eqref}[1]{Equation~(\ref{#1})}
\newcommand{\eqref}[1]{Eq.\,(\ref{#1})}

\newcommand{\eqrngref}[2]{Eqs.\,(\ref{#1}--\ref{#2})}
\newcommand{\figref}[1]{Fig.\,\ref{#1}}
\newcommand{\Figref}[1]{Figure \ref{#1}}
\newcommand{\tabref}[1]{Tab.\,\ref{#1}}
\newcommand{\Tabref}[1]{Table \ref{#1}}
\newcommand{\secref}[1]{\S\ref{#1}}
\newcommand{\Secref}[1]{Section~\ref{#1}}
\newcommand{\p}{\mathcal{P}}

\newcommand{\n}{\mathcal{N}}

\newcommand{\vmu}{\overline{\mu}}
\newcommand{\vsigma}{\overline{\sigma}}
\newcommand{\vlambda}{\overline{\lambda}}
\newcommand{\vI}{\overline{I}}

\begin{document}

\title{Using Bayesian Analysis and Gaussian Processes to Infer Electron Temperature and Density Profiles on the MAST Experiment}

\author{G.~T.~von~Nessi}
\email[]{greg.vonnessi@anu.edu.au}

\author{M.~J.~Hole}
\affiliation{
Research School of Physical Sciences and Engineering,
The Australian National University,
Canberra ACT 0200, Australia
}

\author{the~MAST~Team}
\affiliation{
EURATOM/CCFE Fusion Association, Culham Science Centre,
Abingdon, Oxon, OX14 3DB, UK
}

\begin{abstract}
A unified, Bayesian inference of midplane electron temperature and density profiles using both Thomson scattering (TS) and interferometric data is presented. Beyond the Bayesian nature of the analysis, novel features of the inference are the use of a Gaussian process prior to infer a mollification length-scale of inferred profiles and the use of Gauss-Laguerre quadratures to directly calculate the depolarisation term associated with the TS forward model. Results are presented from an application of the method to data from the high resolution TS system on the Mega-Ampere Spherical Tokamak, along with a comparison to profiles coming from the standard analysis carried out on that system.

\end{abstract}

{52.55.-s, 52.55.Fa}

\maketitle

\section{Introduction}\label{sec::Introduction}
The incoherent Thomson scattering (TS) of laser light off of electrons is the foundation for standard diagnostics used to measure electron temperature temperature and density in fusion devices.\cite{hutchinson2005} Recently, the MAST Thomson scattering (TS) system was upgraded to measuring 130 points across the midplane at a frequency of up to 240Hz.\cite{scannell2008} 

In this work, Bayesian methods are used to infer both $T_e$ and $n_e$ profiles using both TS and interferometric data coming from diagnostics on the Mega-Ampere Spherical Tokamak (MAST). The primary advantage of this analysis is that an absolute calibration of the time-integrated TS laser pulse energy is not required to infer either the thermal electron temperature nor density profiles.

A stochastic model, based on Gaussian processes (GPs)\cite{rasmussen2005}, is used to model the correlation between TS observation points. The noise, signal-variance and length-scale associated with this model are inferred as nuisance parameters in addition to the $T_e$ and $n_e$ profiles. This provides a strong benefit of the presented approach, in that signal noise is strongly decoupled from the inferred profiles; and thus, helps to mitigate the probability of previous errors\cite{cardozo1994, vanmilligen2003, cardozo2003}, where signal noise was given a physical interpretation, from reoccurring. Moreover, the standard perturbative methods for approximating the depolarisation term in the TS forward model, (c.f.   Naito, et. al.\cite{naito1993}) are bypassed in this analysis, in favour of a more direct and flexible numerical integration calculation using Gauss-Laguerre quadratures. Caching of depolarisation calculations is employed to negate the higher computational cost of this approach.

The paper is structured as follows. Section \secref{sec::Bayesian} gives a brief overview of Bayesian inference using diagnostic data and the application of GPs to model the correlation between observation points. Next, the TS and interferometric forward models used are presented with a description of the depolarisation computation. \Secref{sec::Results} present profiles inferred for a high-performance MAST discharge along with a comparison to profiles calculated from the MAST standard analysis. Finally, conclusions and possible extensions are discussed.

\section{Overview of Bayesian Inference}\label{sec::Bayesian}
The objective of any inference scheme is to statistically infer a vector of model parameters, denoted $\vlambda$, given a vector of diagnostic data and associated uncertainties, $\vmu$ and $\vsigma$ respectively. In the Bayesian perspective, inference centres around Bayes' formula:
\begin{equation}
\p(\vlambda|\vmu,\vsigma,\vI)=\frac{\left(\prod_i\p(\mu_i|\vlambda,\sigma_i,\vI)\right)\p(\vlambda)}{\p(\vmu,\vsigma,\vI)},\label{eq1}
\end{equation}
where $\vI$ denotes background assumptions. To keep notation uncluttered, $\vI$ is dropped for the rest of the paper, with background assumptions being explicitly indicated where appropriate. The parts of \eqref{eq1} and their application/interpretation in the context of diagnostic data are well documented in the literature\cite{vonnessi2012,svensson2008,sivia2006,jaynes2003} and will not be discussed in detail here.

TS and interferometer uncertainties are assumed to be pairwise uncorrelated, with each observation having an associated likelihood of the assumed form
\begin{equation}
\p(\mu_i|\vlambda,\sigma)=\n(\mu-\mathcal{F}(\vlambda),\sigma^2),\label{eq2a}
\end{equation}
where $\n(\mu,\sigma^2)$ is a Gaussian distribution of mean $\mu$ and $\sigma^2$ the variance; $\mathcal{F}(\vlambda)$ is the forward model associated with the given diagnostic. Justifications for this form of the likelihood are given elsewhere \cite{vonnessi2012, sivia2006}.

A GP is employed to define the covariance matrix reflecting a spatial correlation between TS observation points:
\begin{equation}
K_{ijk} := \zeta_{k}^2\exp\left(-\frac{(R_i-R_j)^2}{2\tau_{k}^2}\right)+\eta_{k}^2\delta_{ij},\label{eq3}
\end{equation}
where $R_i$ and $R_j$ represent the radial coordinate of the $i$th and $j$th TS observation point along the midplane; and the $k$ index indicates correspondence to electron temperature or density measurements, with a value of $0$ or $1$ respectively. In \eqref{eq3} $\zeta_k$, $\tau_k$ and $\eta_k$ are commonly referred to as hyper-parameters\cite{rasmussen2005}: non-physical quantities which help characterise the prior and/or likelihood in the overall inference. In this analysis, these are scalar quantities uniquely associated with each profile: $\tau_k$ reflects the average radial length-scale over which the profile is changing; $\sigma_k$ is the signal variance, which serves to decouple the average profile gradient from the length-scale; and $\eta_k$ is the average scalar noise on the profile. With the expression in \eqref{eq3}, the prior for the inference is proportional to
\begin{eqnarray}
\p(\vlambda) & = & \p(\vec{T_e},\vec{n_e}, \tau_0, \tau_1, \zeta_0, \zeta_1, \eta_0, \eta_1)\nonumber\\
&  \propto & \n\left(\vec{T_e}^TK_{0}^{-1}\vec{T_e}\right)\n\left(\vec{n_e}^TK_{1}^{-1}\vec{n_e}\right)\prod_{k=0}^1\left(\mathbf{1}_{[0,10]}(\vec{\tau_k})\mathbf{1}_{[0,10]}(\vec{\zeta_k})\mathbf{1}_{[0,10]}(\vec{\eta_k})\right),\label{eq4}
\end{eqnarray}
where $\n$ are zero-mean Gaussian distribution with their covariance matrix shown in the argument, and $\mathbf{1}_{[a,b]}(x)$ indicates a uniform distribution of the variable $x$ over the closed interval $[a,b]$. The upper bounds for the $\tau_k, \zeta_k$ and $\eta_k$ uniform distributions were empirically selected so as to not preclude any physically attainable profiles on MAST, as determined by analysing 36 different shot/time slices.

\Eqref{eq4} shows the prior as a product of uniform and zero-mean Gaussian distributions. As the covariance matrices for the Gaussians over $T_e$ and $n_e$ are themselves characterised by \emph{inferred} hyper-parameters, the prior in \eqref{eq4} embodies a stochastic model of the spatial correlation between different TS observation points. For a fixed set of hyper-parameters, one can think of this prior as favouring $T_e$ and $n_e$ profiles convolved with Gaussian kernels (i.e. mollifications\cite{evans2010}) of fixed widths corresponding to $\tau_k$ in the inference. As the hyper-parameters themselves are inferred with minimal constraint, the analysis is also able to infer the average length-scale, signal variance and noise variance for both the $T_e$ and $n_e$ profiles, as intrinsically held by the data. Further details on these points can be found in Rasmussen\cite{rasmussen2005}.

\section{Forward Model}\label{sec::FM}
In the MAST TS system, Thomson scattered light from each observation point is spectrally divided into four bands via a polychrometer. Each filtered band is then focused onto an avalanche photodiode (APD), translating the integrated intensity over the spectral band into a voltage signal. When integrated over the TS laser pulse length, a quantity is produced which is sensitive to both the thermal electron temperature and density at the associated observation point:
\begin{eqnarray}
V_{TS} = C_Sn_eE_L\int\frac{\phi(\lambda)}{\phi(\lambda_L)}\frac{S(\lambda_s, \lambda_L, \theta, T_e)}{\lambda_L}\,d\lambda,\label{eq::ts3}
\end{eqnarray}
where $C_S$ encompasses a collection of known, fixed system constants; $E_L$ is the integrated laser energy; $\phi(\lambda)$ is the polychrometer response functions; $\lambda_L$ is the TS laser wavelength; $\lambda_s$ is the wavelength of the scattered photons; $\theta$ is the scattering angle; and $S(\lambda_s, \lambda_L, \theta, T_e)$ the standard Selden expression. \cite{scannell2007, scannell2008} The Selden relation relates the intensity of scattered light off of thermal electrons to $T_e$ and is standard in the literature; but the relation is recalled here for completeness (note the variable transformations in \eqrngref{eq::ts4d}{eq::ts4e}):
\begin{eqnarray}
S(\epsilon,\theta,2\alpha) & = & S_Z(\epsilon,\theta,2\alpha)q(\epsilon,\theta,2\alpha),\label{eq::ts4a}\\
S_Z(\epsilon,\theta,2\alpha) & = & \frac{\exp(-2\alpha x)}{2K_2(2\alpha)(1+\epsilon)^3}\left[2(1-\cos\theta)(1+\epsilon)+\epsilon^2\right]^{-1/2},\label{eq::ts4b}\\
q(\epsilon,\theta,2\alpha) & = & 1+\frac{2x}{y}\exp(2\alpha x)\left(y^2\int_x^\infty\frac{\exp(-2\alpha\xi)}{(\xi^2+u^2)^{3/2}}\,d\xi-3\int_x^\infty\frac{\exp(-2\alpha\xi)}{(\xi^2+u^2)^{5/2}}\,d\xi\right),\label{eq::ts4c}\\
\epsilon & := & \frac{\lambda_s-\lambda_i}{\lambda_i},\quad 2\alpha := \frac{m_ec^2}{T_e},\quad u=\frac{\sin\theta}{1-\cos\theta},\label{eq::ts4d}\\
x & := & \left(1+\frac{\epsilon^2}{2(1-\cos\theta)(1+\epsilon)}\right)^{1/2},\quad y := \frac{1}{(x^2+u^2)^{1/2}},\label{eq::ts4e}
\end{eqnarray}
where $q(\epsilon,\theta,2\alpha)$ represents the relativistic depolarisation correction term.\cite{naito1993,scannell2007,scannell2008}

For TS systems, the expressions presented by Naito, et. al.\cite{naito1993} are normally used to approximate $q(\epsilon,\theta,2\alpha)$. In this work, however, Gauss-Laguerre quadratures are utilised to provide a more direct and flexible calculation of the depolarisation correction. The fundamental difference in this approach is that it is a non-perturbative calculation, which contrasts the approximations of Naito, et. al.\cite{naito1993} that are based Taylor expansions. Indeed, changing variables according to $v = 2\alpha(\xi-x)$ in \eqref{eq::ts4c} gives
\begin{equation}
q(\epsilon,\theta,2\alpha) = 1+\frac{4\alpha x}{y}\left(y^2\int_0^\infty\frac{\exp(-v)}{\left(\left(\frac{v}{2\alpha}+x\right)^2+u^2\right)^{3/2}}\,dv-3\int_0^\infty\frac{\exp(-v)}{\left(\left(\frac{v}{2\alpha}+x\right)^2+u^2\right)^{5/2}}\,dv\right),\label{eq::ts5}
\end{equation}
which can be integrated directly using a Gauss-Laguerre quadrature\cite{hildebrand1987}. The initial computational cost of the quadrature construction is offset by the fact that the integrals in \eqref{eq::ts5} are computed via pre-calculated quadrature poles and weights. Gauss-Laguerre quadratures are very accurate/efficient for calculating integrals of the form seen in \eqref{eq::ts5} and allow for arbitrary levels of accuracy to be specified by simply changing the number of quadrature points.\cite{hildebrand1987}

The primary issue with a forward model based on \eqref{eq::ts3} is that it requires an absolute calibration of $E_L$, in addition to the calibrations reflected in the value of $C_S$. Fortunately, MAST has a midplane CO$_2$ interferometer that provides a line-integrated measurement of $n_e$ in the midplane. Using this data, a simple coordinate transformation is employed to construct the integrated electron density along the TS laser's line of sight (also in the midplane). As $E_L$ is the same for all TS observation points, it can be inferred as a nuisance parameter (i.e. an unphysical parameter which is integrated out in the inference of physical model parameters) that can also absorb any error in $C_S$, with $n_e$ still being well-constrained by both TS and the line-integrated interferometeric observation. This serves to greatly reduce the errors seen on the inferred profiles, even when $E_L$ is given a physically unconstraining uniform prior, as in this analysis (see \secref{sec::Results}).

\section{Results}\label{sec::Results}
To demonstrate the analysis detailed above, an inference and subsequent comparison is made against profiles coming from the standard TS analysis carried out on MAST. This comparison is made on discharge 24600 at 280ms, which is a L-mode discharge in a DnD configuration with 3.35MW of co-injected NBI heating. The Bayesian inference was carried out using a 1000 point Gauss-Laguerre quadrature to calculate $q(\epsilon,\theta,2\alpha)$, with posterior moments taken from sampling statistics obtained via a specialised implementation of Skilling's nested sampling (NS) algorithm detailed elsewhere\cite{vonnessi2012b} (see Sivia and Skilling\cite{sivia2006} for details on NS). The sampling results were also independently validated by comparison with samples generated from a Hamiltonian Markov Chain Monte Carlo (HMCMC) algorithm (see MacKay\cite{mackay2003} for details on HMCMC).

\begin{figure}[!htb]
\includegraphics[width=\textwidth]{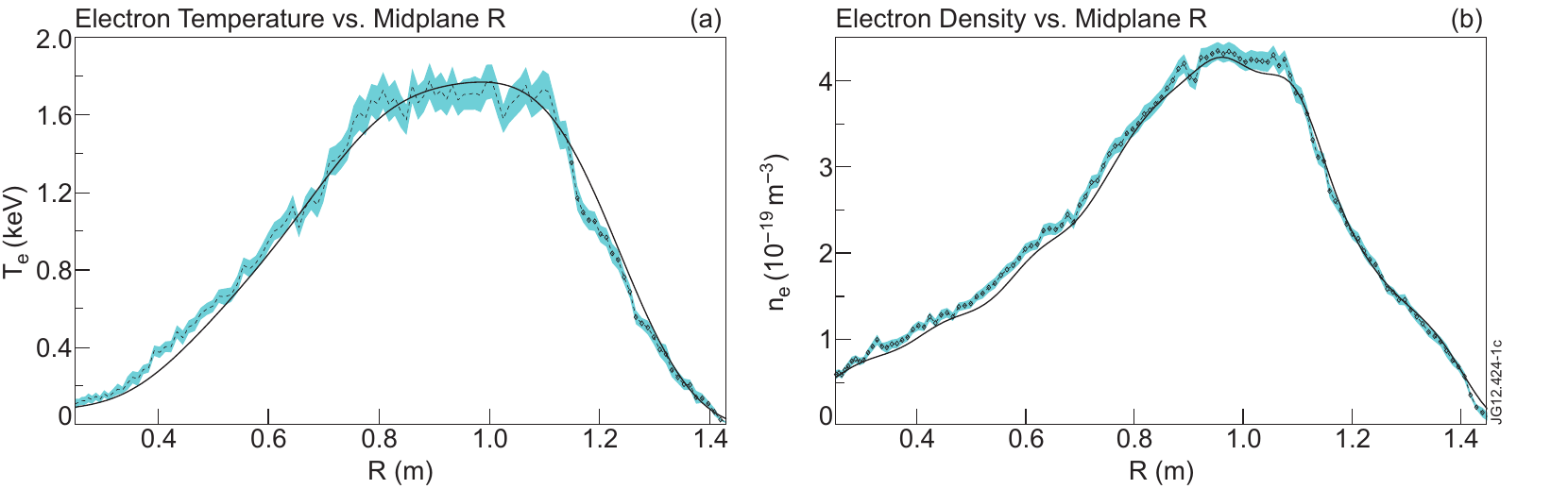}  
\caption{\label{fig::fig1} Comparison of inferred $T_e$ and $n_e$ profiles inferred from TS data for shot \#24600 at 280ms, with $T_e$ and $n_e$ profiles corresponding to (a) and (b) respectively. The heavy lines indicate profiles inferred using the Bayesian analysis described in \secref{sec::Bayesian}. The dashed lines marked with diamonds and coloured region reflect MAST scheduler output for the corresponding profiles and uncertainty respectively. Uncertainties associated with the expectation of the Bayesian inferred profiles are too small to resolve visually and have been suppressed from both figures.}
\end{figure}

\Figref{fig::fig1} shows that the inferred, expectation of the profiles show generally, very good agreement with the profiles coming from standard analysis. The most striking difference between both sets of profiles is the difference in uncertainties. Indeed, the nature of the unified Bayesian inference indicates that the profiles are very well constrained. This can be understood by noting that $T_e$ and $n_e$ are being inferred from four APD measurements (nominally) at every observation point, which are in addition to the global constraints provided by interferometry and the prior, rendering the inference as being strongly over-determined. The presence of global constraints and over-constraining local observations make the inference very robust against APD errors, mis-calibrations or signal loss. In contrast, the MAST analysis uncertainty reflects a maximum entropy result (i.e. with minimal prior assumptions relative to the forward model used) and is expected to have higher uncertainties than results utilising the prior in \eqref{eq4} (or any analogy thereof).

As mentioned in \secref{sec::Bayesian}, this inference yields mollified profiles, along with the associated hyper-parameters represented in \eqref{eq3}. Thus, the results produced have a different interpretation (i.e. they need to be understood in relation to their respective hyper-parameters) than the profiles from the standard MAST analysis. Indeed, \figref{fig::fig1} can only be viewed as a sanity check for the global shape/scale of the $T_e$/$n_e$ profiles, as the fine-scale structure information coming from the Bayesian analysis is largely contained in the inferred hyper-parameters. In this context, it is not surprising to see the inferred profiles lying outside the errors bars associated with the MAST analysis. One of the main advantages for such a comparison, is that one can qualitatively see how much of the profile structure can be attributed to a scalar noise term. Quantitatively, one can compare the scalar noise term, $\eta_{k}$, of the profile to the random errors modelled for the diagnostic to validate uncertainties produced by explicitly propagating error through the forward model itself. Finally, one may adjust the prior on any of the hyper-parameters in the inference to reflect a priori knowledge of the length-scales or random noise seen in the system; this can even be taken to the point where these hyper-parameters are even fixed at particular values. Of course, such priors will effect the structure of the inferred profiles; e.g. fixing a relatively small length-scale and scalar noise will yield profiles which have a structure closer to those produced by the MAST analysis. Exploration of physics using these more informed priors on the hyper-parameters is a current research endeavour.

\begin{table}[!hbt]
\centering
\begin{tabular}{|c|c|}
\hline
\parbox{.35\textwidth}{\textbf{Nuisance Parameter (Unit)}} & \parbox{.6\textwidth}{\textbf{Sampling Expectation w/ 95\% confidence interval}}\\
\hline
$E_L\ (J)$ & $(1.345\times10^{-1})^{+1.703\times 10^{-3}}_{-1.561\times 10^{-3}}$\\
$\tau_{0}\ (m)$ & $(2.296\times10^{-1})^{+6.655\times 10^{-3}}_{-8.101\times 10^{-3}}$\\
$\zeta_{0}$ & $1.006^{+2.822\times10^{-3}}_{-5.427\times10^{-3}}$\\
$\eta_{0}$ & $(6.698\times10^{-3})^{+3.209\times10^{-3}}_{-6.073\times10^{-3}}$\\
$\tau_{1}\ (m)$ & $(9.974\times10^{-2})^{+1.761\times10^{-3}}_{-1.081\times10^{-2}}$\\
$\zeta_{1}$ & $1.511^{+3.039\times10^{-3}}_{-2.839\times10^{-3}}$\\
$\eta_{1}$ & $(6.650\times10^{-3})^{+5.353\times10^{-3}}_{-5.580\times10^{-3}}$\\
 \hline
\end{tabular}
\caption{Table of sampling statistics of nuisance parameters used in the inference of $T_e$ an $n_e$ profiles from TS data for discharge \#24600 at 280ms.}
\label{tab::nuissanceParameters}
\end{table}

\Tabref{tab::nuissanceParameters} shows sampling expectations and uncertainties associated with the nuisance parameters presented in \eqref{eq3} and \eqref{eq::ts3}. While the values in \tabref{tab::nuissanceParameters} are treated as nuisance parameters, they have been included to give the reader some context for the length scales, signal and noise variance that are typically encountered with MAST TS data. Again, the uncertainties on these values are small relative to the inferred values, as the inference is over-constrained to the point where uncertainties due to degeneracies in highly likely model parameter configurations are all but eliminated. 

Finally, as the inference outlined in this paper is non-analytic in nature, having approximately 270 model parameters for a given discharge, it is slower than the standard analysis. Indeed, to calculate statistical moments of the profiles using NS takes approximately thirty minutes per inference (on average), on a 2.2GHz processor with 8GB of memory. It is a current research focus to speed up this inference by developing a parallelised version of sampling algorithm.

\section{Conclusions}\label{sec::Conclusions}
A new method for the unified Bayesian inference of thermal electron temperature and density profiles has been demonstrated, which is also able to infer length-scales and scalar noise parameters intrinsically contained within the diagnostic data. By employing GPs, the average length-scale, signal variance and noise variance are inferred as nuisance parameters, which directly yield profiles where noise is minimised. Finally, a new approach to calculating the depolarisation correction in the TS forward model is presented, which is simple to implement and affords easy adjustment of calculation accuracy.

\begin{acknowledgments}
This work was jointly funded by the Australian Government through International Science Linkages Grant CG130047, the Australian National University, the RCUK Energy Programme under grant EP/I501045 and the European Communities under the contract of Association between EURATOM and CCFE. The views and opinions expressed herein do not necessarily reflect those of the European Commission.
\end{acknowledgments}

\bibliographystyle{unsrt}
\bibliography{tsGPRefs}

\end{document}